\documentclass[prl,twocolumn,showpacs,showkeys,amsmath,amssymb]{revtex4-1}
\usepackage{graphicx}
\usepackage{dcolumn}
\usepackage[sort&compress]{natbib}
\usepackage{subfigure}
\usepackage{ifpdf}
\usepackage[version=3]{mhchem}
\usepackage{bm}

\begin{document}

\title{Mapping electronic reconstruction at the metal/insulator interfaces in \ce{LaVO_3/SrVO_3} heterostructures}

\author{Haiyan Tan}
\email{Haiyan.Tan@cemes.fr}
\affiliation{CEMES-CNRS, Univ. de Toulouse, nMat group, BP94347, 31055, Toulouse Cedex 4, France}
\affiliation{EMAT, University of Antwerp, Groenenborgerlaan 171, B-2020 Antwerpen, Belgium}

\author{Ricardo Egoavil}
\author{Armand B\'ech\'e}
\author{Gerardo T Martinez}
\author{Sandra Van Aert}
\author{Jo Verbeeck}
\author{Gustaaf Van Tendeloo}
\affiliation{EMAT, University of Antwerp, Groenenborgerlaan 171, B-2020 Antwerpen, Belgium}

\author{H\'el\`ene Rotella}
\author{Philippe Boullay}
\author{Alain Pautrat}
\author{Wilfrid Prellier}
\affiliation{CRISMAT, ENSICAEN, CNRS UMR 6508, 6 Boulevard Mar\'echal Juin, F-14050 Caen Cedex 4, France}

\date{\today}

\begin{abstract}

A \ce{(LaVO_3)_6/(SrVO_3)_3} superlattice is studied with a combination of sub-{\AA} resolved scanning transmission electron microscopy and monochromated electron energy-loss spectroscopy. The V oxidation state is mapped with atomic spatial resolution enabling to investigate electronic reconstruction at the \ce{LaVO_3}/\ce{SrVO_3} interfaces. Surprisingly, asymmetric charge distribution is found at adjacent chemically symmetric interfaces. The local structure is proposed and simulated with double channeling calculation which agrees qualitatively with our experiment. We demonstrate that local strain asymmetry is the likely cause of the electronic asymmetry of the interfaces. The electronic reconstruction at the interfaces extends much further than the chemical composition, varying from 0.5 to 1.2~nm. This distance corresponds to the length of charge transfer previously found in the \ce{(LaVO_3)_m}/\ce{(SrVO_3)_n} metal/insulating and the \ce{(LaAlO_3)_m}/\ce{(SrTiO_3)_n} insulating/insulating interfaces.

\end{abstract}


\maketitle

\section{Introduction}

Transition metal oxides (TMO) derived from the perovskite structure receive considerable attention because of their fascinating transport and magnetic properties that can be tuned by cationic substitution on the A and/or B-site but also by the introduction of vacancies on both cation and anion sublattice. When deposited in the form of epitaxial thin films, these properties can be further tailored by the effect of the substrate strain. In case of artificial superlattices (SLs), the introduction of oxide interfaces between different materials \cite{Ohtomo2002, Ohtomo2004} gave rise to materials where electronic and/or magnetic properties are governed by exotic states present at these interfaces. As an example, magnetism, conductivity and superconductivity observed at the interface between insulating non magnetic oxides \ce{LaAlO_3} (LAO) and \ce{SrTiO_3} (STO), are believed to be closely related to the interface electronic structure and film thickness \cite{Ohtomo2002, Wang2003b, Ohtomo2004, Nakagawa2006, Reyren2007, Kourkoutis2007a, Brinkman2007}.

The physics behind these materials involves many competing parameters that complicate the interpretation of the observed behaviour often leading to controversy \cite{Millis2011}. Understanding the properties of such heterostructures needs advanced methods to probe chemical, structural and electronic changes at the oxide interfaces \cite{Hwang2012,Monkman2012}. Scanning transmission electron microscopy (STEM) combined with electron energy loss spectroscopy (EELS) is one of these state-of-the-art techniques capable to investigate both atomic and electronic reconstruction at complex oxide interfaces \cite{Ohtomo2002,Ohtomo2004, Nakagawa2006,Cantoni2012, Turner2012a, Vantendeloo2012}.

Here we focus on one high quality \ce{(LaVO_3)_m/(SrVO_3)_n} SL presenting interfaces between the antiferromagnetic Mott-Hubbard insulator \ce{LaVO_3} \ce{(LVO)} and the paramagnetic metal \ce{SrVO_3} \ce{(SVO)}. In bulk materials, vanadium adopts a 3+ state in LVO, a 4+ state in SVO and, consequently, a mixed 3+/4+ oxidation state in the La$_{1-x}$Sr$_x$VO$_3$ solid solution, leading to critical behavior of a metal-insulator transition depending on the doping level $x$ \cite{Imada1998, Miyasaka2000}. In \ce{(LVO)_m/(SVO)_n} SLs \cite{Sheets2007, Sheets2009}, by varying the m and n values, we can artificially create a large variety of heterostructures yielding a mixed V$^{3+}$/V$^{4+}$ oxidation state which, contrary to the bulk materials, is expected to be localized at the interfaces between LVO and SVO layers. Like the LAO/STO model system, the LVO/SVO interfaces shall exhibit a ``polar catastrophe" scenario \cite{Nakagawa2006} requiring a charge transfer of half an electron per \ce{VO_2} unit. The existence of vanadium in a mixed valence state in adjacent \ce{VO_2} sheets with the possibility of partially localized electron holes at interfaces and how this affects the transport and magnetic properties \cite{Sheets2007,Sheets2009,Luders2009,David2011} remains an important issue in such heterostructures. Carrier localization effects at the interfaces have been partially confirmed by optical spectrocopy \cite{Jeong2011} but could not, at that time, be inferred from STEM and EELS due to both atomic-scale disorders at the interfaces and limited resolution of the STEM-EELS experiments \cite{Boullay2011}. In the present study, with the improved spatial and energy resolution of modern STEM EELS instruments, we have obtained simultaneously at atomic resolution an elemental and, more importantly, a V oxidation state map enabling us to evaluate charge transfer at the LVO/SVO interfaces in a quantitative way.

\section{Experiments}

A \ce{(LVO)_6/(SVO)_3} SL was prepared by pulsed-laser deposition on (100)-\ce{SrTiO_3} substrates as described in details in ref.\cite{Sheets2007}. High-angle annular dark-field (HAADF) STEM and STEM-EELS experiments were performed on the Qu-Ant-EM microscope at the University of Antwerp. The microscope consists of a FEI Titan 80-300 ``cubed" microscope fitted with a double aberration-corrector for both probe-forming and imaging lenses, a monochromator and a GIF Quantum energy filter\cite{Gubbens2010}. The microscope was operated at 120~kV acceleration voltage to avoid knock-on damage\cite{Egerton2004, Botton2010}. The convergence and collection angles for EELS acquisition are 18 and 120~mrad, respectively. The inner and outer angle for the HAADF detector were 70 and 160~mrad. A Fourier transform of the acquired HAADF-STEM images indicates information transfer down to 0.8~{\AA}. The Wien-filter monochromator was excited and provides an energy resolution of 300~meV at the \ce{V-L_{2,3}} core loss edge for the LVO and SVO single thin films. To cover both the \ce{V-L_{2,3}}(513~eV) and \ce{La-M_{4,5}} edge (832~eV) in the same spectrum, a 0.25~eV/pixel dispersion was chosen for the superlattice sample. This slightly reduces the energy resolution \cite{Gubbens2010} but the fingerprints remains clear. Low beam intensity ($\sim$20~pA), pixel sampling (0.70~{\AA}/pixel) and dwell time (0.2~s/pixel) were chosen as a compromise between EELS signal-to-noise ratio, beam damage and sample drift \cite{Tan2011}. A HAADF image was acquired simultaneously with the EELS experiment using software synchronized acquisition. This image confirms minimum sample drift and no significant beam damage during the acquisition.

HAADF-STEM images were quantified by means of statistical parameter estimation theory following the approach presented in \cite{Aert2009}. The lattice displacement was measured as the distance between neighbouring La/Sr columns computed from the estimated atomic column positions. The lattice spacings are compared to the substrate and are expressed as relative strain values.

For the averaged elemental profile, no data pre-treatment was applied except summing the spectra along the interface direction. For the 2D valence maps, the raw spectrum image was smoothed with a Gaussian low pass filter (FWHM=0.12~nm) to improve the signal-to-noise ratio. The data were then background subtracted by fitting a commonly-used power-law model $AE^{-r}$. A 100~eV region before the V-L$_{2,3}$ edge was used to fit the power law parameters $A$ and $r$ for each pixel. The raw elemental maps of vanadium shows that the vanadium layers grow continuously as expected. Sample thickness and probe current variations are normalised by assuming a constant V signal averaged over a unit cell.

\section{Result and discussion}

\begin{figure}
\centering
\includegraphics[width=0.99\linewidth]{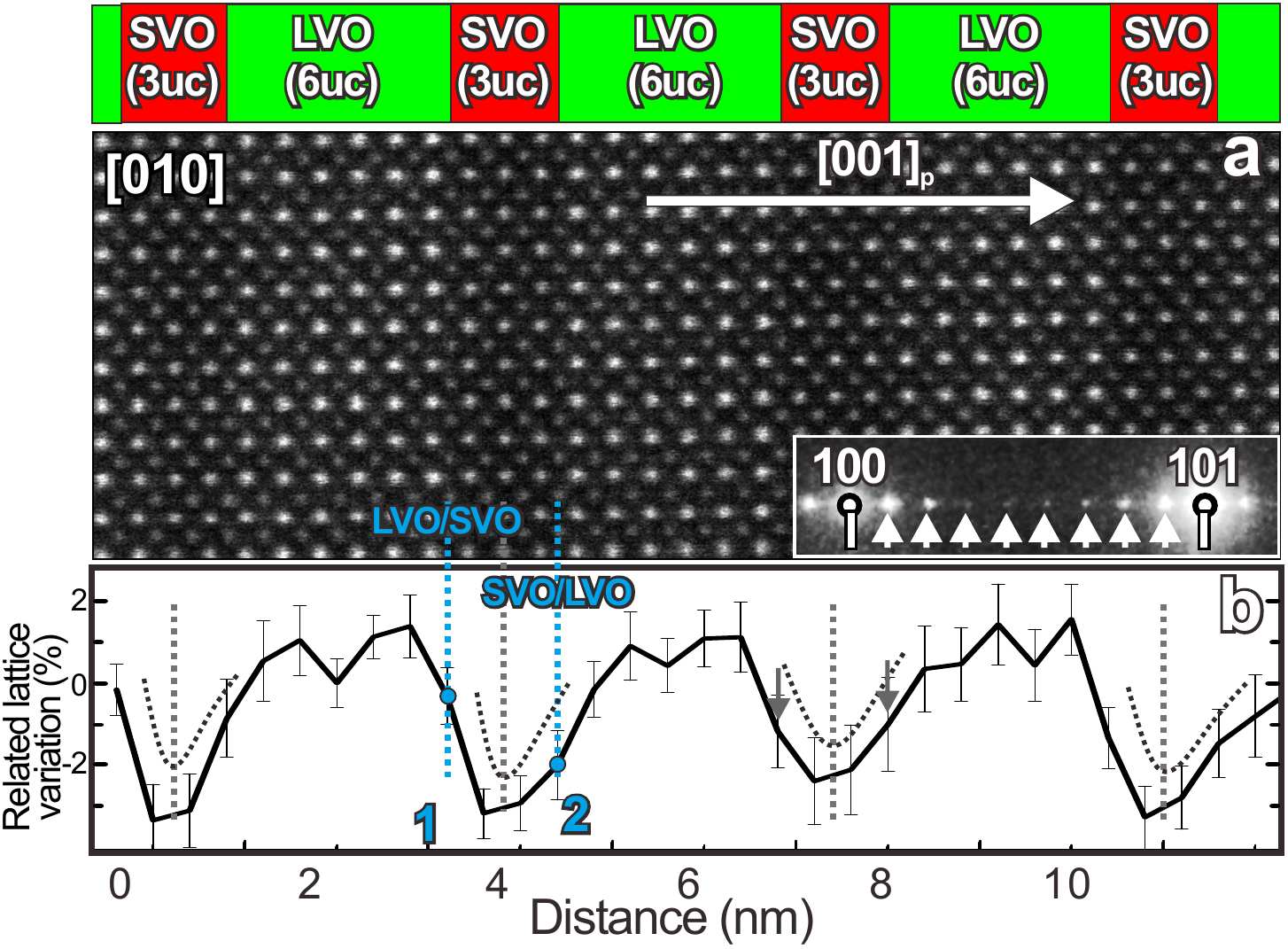}
\caption{(a) HAADF-STEM image of the \ce{(LaVO_3)_6/(SrVO_3)_3} SL with the film structure sketched on the top. The inset shows part of the diffraction pattern with satellite spots corresponding to the 9 perovskite unit cells superlattice structure. (b) Lattice displacement profile (black) shows the mean displacements along the [001] direction expressed in terms of deviation in percentage from the STO substrate. The error bars show confidence intervals for the mean values obtained by averaging along the [010] direction. Asymmetric strain is observed between LVO/SVO and SVO/LVO interfaces. Dashed lines serve as guide to the eye.}
\label{Overview}
\end{figure}

Fig.\ref{Overview}a shows a HAADF-STEM image of the \ce{(LVO)_6/(SVO)_3} film with excellent spatial resolution. The brighter/darker layers are corresponding to the LVO/SVO layers respectively as expected from the Z-contrast property of HAADF imaging \cite{Hartel1996}. The straight interfaces and the sharp contrast between the layers illustrates that the boundaries are atomically flat. Its electron diffraction pattern (Fig.\ref{Overview}a inset) exhibits main reflection spots corresponding to the pseudocubic perovskite subcell (a=3.90(4)~{\AA}). Sharp satellite spots are also observed along the [001]$_p$ growth direction, indicating a good quality SL over the whole thickness of the film (Fig.\ref{Overview}a inset). The superlattice periodicity is 9 times larger than the perovskite subcell as expected for a \ce{(LVO)_6/(SVO)_3} SL.

Since the lattice sizes of LVO and SVO are originally different in bulk materials \cite{Sheets2007}, the c lattice parameter of LVO is expanded (1.5(1.0)\%) whereas for SVO it is contracted (-3(1)\%) with respect to the c parameter in the STO substrate (Fig.\ref{Overview}b black). The lattice displacement variation along the growth direction shows an asymmetrical trend on average. The contraction is more abrupt at interface 1 of SVO layer and more gradual at interface 2. This can be understood since the strain acts over a long range up to 50~nm \cite{Gariglio2007}. The interface symmetry of the SL is broken by the substrate on one side and vacuum on the other side. This asymmetric lattice parameter variation indicates an asymmetric strain situation between interfaces 1 and 2. The asymmetric strain will impact the local electronic structure at the interfaces as we will show next. \cite{Vailionis2011, Koster2012}. Also note that the strain varies at different interfaces and occasionally the asymmetry may switch as indicated by grey arrows in Fig.\ref{Overview}b.

\begin{figure}
\centering
\includegraphics[width=0.99\linewidth]{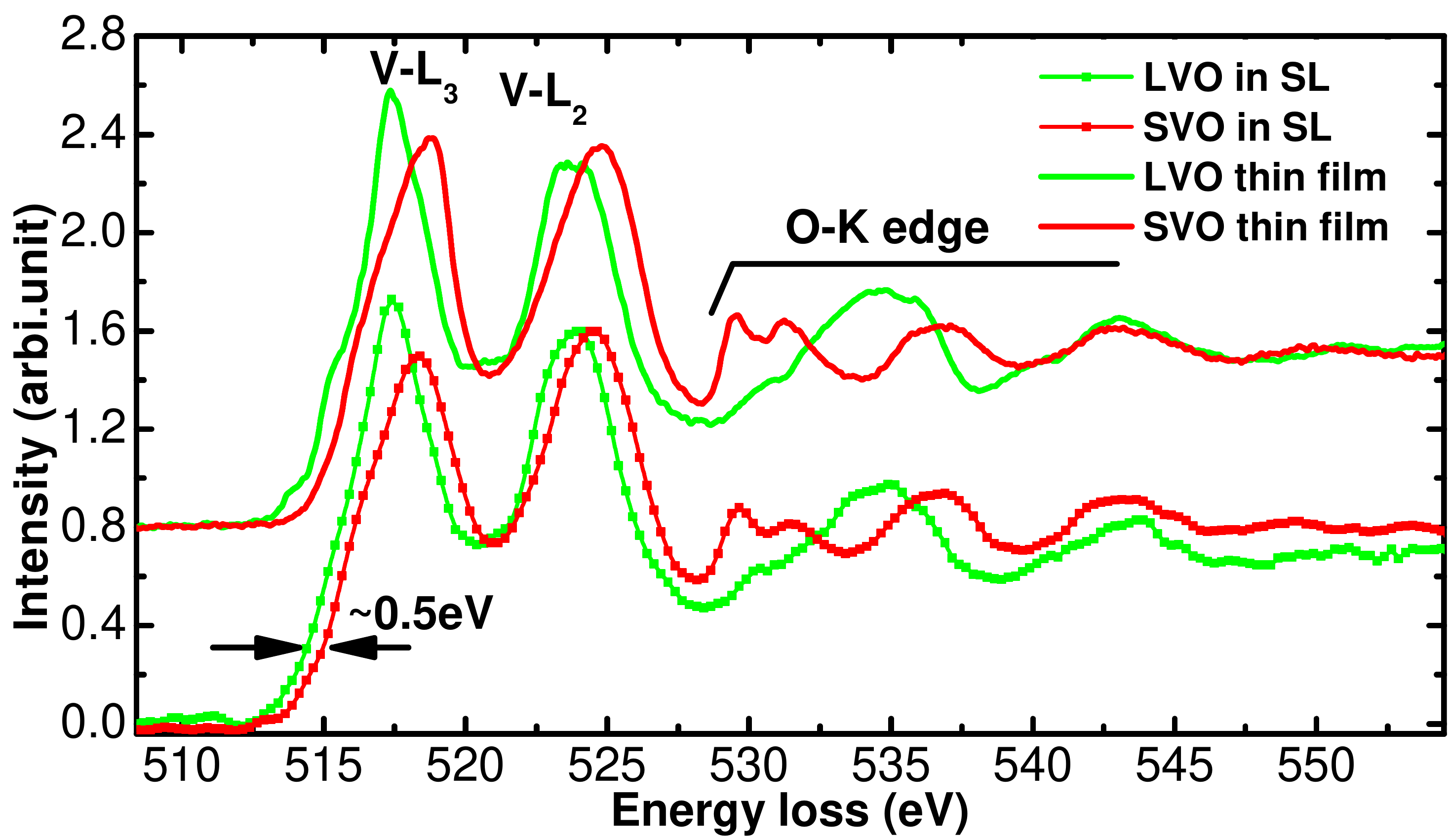}
\caption{LVO and SVO spectra from both the \ce{(LaVO_3)} and \ce{(SrVO_3)} reference thin films and their signal in the SL \ce{(LaVO_3)_6}/\ce{(SrVO_3)_3}. For SLs, a weighted LVO spectrum is subtracted from the SVO spectrum to eliminate the delocalized LVO signal in the SVO layer.}
\label{Spectrum}
\end{figure}

The straight interfaces and the sharp contrast between the layers illustrates that both the LVO/SVO and SVO/LVO interfaces are atomically flat (Fig.\ref{Overview}a). In order to probe how vanadium changes its oxidation state across the interface, the electron configuration of V (\ce{V^{3+}}~(3\ce{d^2}) and \ce{V^{4+}}~(3\ce{d^1})) is revealed from the fine structures of its \ce{L_{2,3}} EELS core-loss edge \cite{Tan2012, Garvie1994, Kourkoutis2006}. Fig.\ref{Spectrum} (top) shows spectra from a single layer LVO and SVO thin film ($\sim$100~nm), which were also prepared in the same way as the SL. Their fine structure agrees well with the bulk spectra from literature with slightly more details because of our better energy resolution (0.3~eV)\cite{Kourkoutis2006}. This confirms that the V in the grown layers has the same chemical environment as the corresponding bulk materials and oxygen deficiency effects could not be detected within experimental capability. The higher peak onset ($\approx$0.5~eV) of the V \ce{L_{2,3}} edge for the SVO film compared to the LVO film indicates a higher oxidation in SVO as expected \cite{Kourkoutis2006, Tan2012}.

Spectra of SVO and LVO in the SL can be obtained from the center of the SVO and LVO layers in the SL. Here we derive the \ce{V^{4+}} signal from the SVO part of the SL by subtracting a scaled LVO signal so that the La signal becomes zero. This should create that part of the spectrum which is truly coming from SVO only without spurious signal from the neighboring LVO layers. The LVO part of the SL being twice thicker, contribution from SVO to the spectrum recorded at the center of the LVO layers is assumed to be negligible. Fig.\ref{Spectrum} (bottom) shows the averaged spectra of LVO and SVO after subtracting the intermixing with a slightly lower energy resolution of approximately 0.5~eV. The fine structure of the V \ce{L_{2,3}} edges of the LVO and SVO layers in the SL appear almost identical with that of our LVO and SVO thin film references. Their O-K edges are also similar which implies that any oxygen deficiency differences are also negligible in the SL. We use the reference spectra from the single thin films as fingerprints for the V \ce{L_{2,3}} edges in either \ce{V^{3+}} and \ce{V^{4+}} oxidation state. The V \ce{L_{2,3}} edge at any position in the LVO/SVO SL is then fitted with a linear combination of these reference spectra in order to extract the contribution from the \ce{V^{3+}} and \ce{V^{4+}} states \cite{Ohtomo2002, Kourkoutis2006, Garcia2010, Tan2012}.

\begin{figure}
\centering
\includegraphics[width=0.99\linewidth]{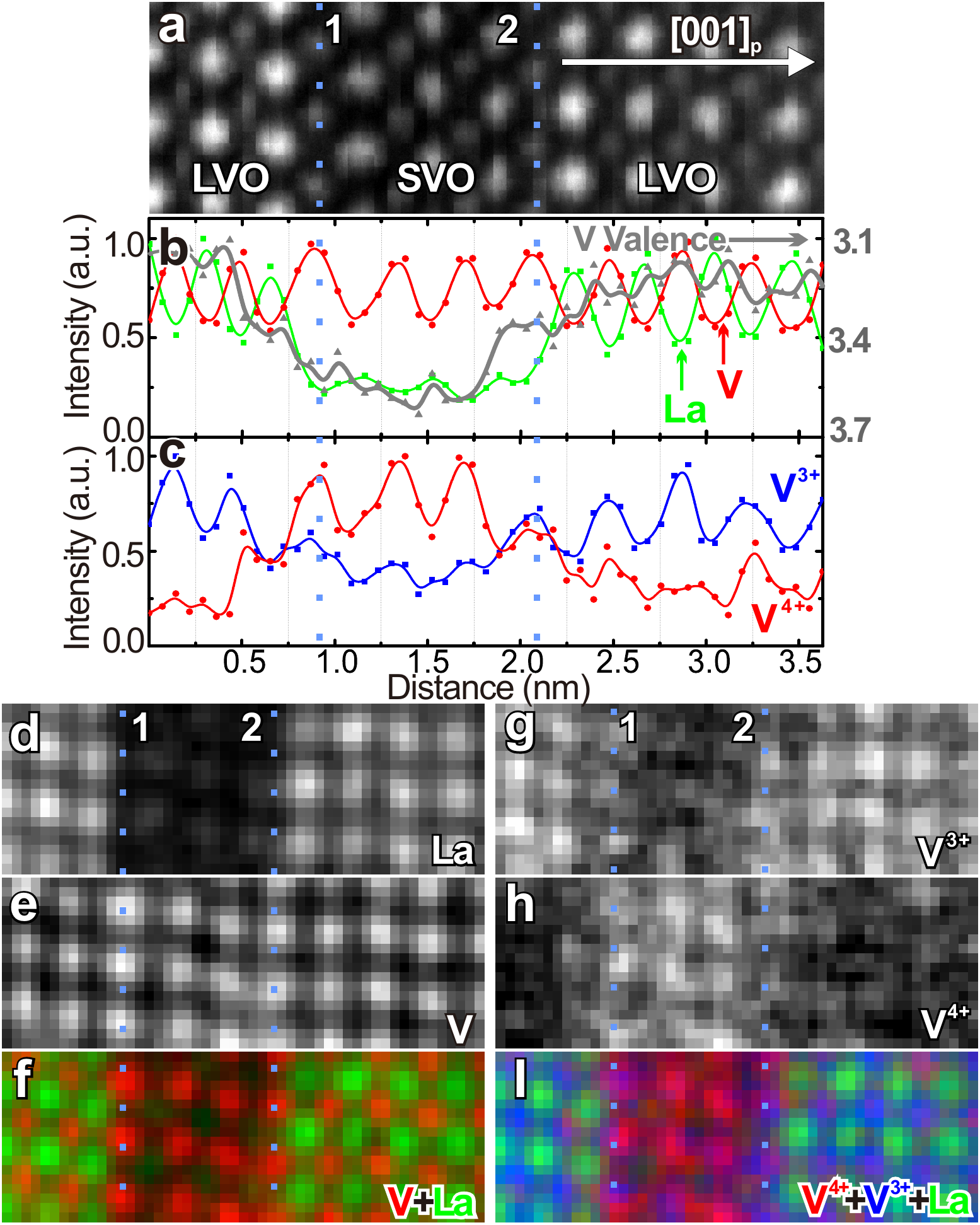}
\caption{(a) dark field image acquired simultaneously with its EELS spectral data from this region. (b)profiles of La signal (green), V signal (red) and V oxidation state (gray) averaged along the interface direction. (c) profiles of \ce{V^{3+}} and \ce{V^{4+}} signal showing the asymmetry at interface 1 and 2. (d)(e) elemental color map of V and La, respectively, and their combined color map in (f). (g)(h) spectral weight maps of \ce{V^{3+}} and \ce{V^{4+}} signals, respectively. (i) combined color map with \ce{V^{4+}} (red), \ce{V^{3+}} (blue) and La (green).}
\label{Vmap}
\end{figure}

Fig.\ref{Vmap}a shows a dark field image acquired simultaneously with the EELS spectrum image at each position containing the \ce{V-L_{2,3}}(513~eV), \ce{O-K}(532~eV) and \ce{La-M_{4,5}}(832~eV) excitation edges in one spectrum. From low loss EELS we estimate the sample to be 10$\sim$15~nm thick meaning that we have approximately 40 atoms in projection. Fig.\ref{Vmap}d and Fig.\ref{Vmap}e show the elemental maps of La and V after data treatment. The combined elemental color map (Fig.\ref{Vmap}f) shows the atomic SL structure with the interfacial V layer at interface 1 and 2 indicated. The profiles of V and La were averaged along the interfaces and shown in Fig.\ref{Vmap}b. Since both La and Sr occupy the same atomic sites, their information can be considered complementary and a chemical diffusion map indicating La/Sr intermixing can be estimated from the normalised La intensity (Fig.\ref{Vmap}b). In the La profile, the valley at the atomic layers 1 and 2 has the same depth and the peak height of the neighboring (left of 1, right of 2) LaO layers are almost equal. Qualitatively, the interfaces can be considered as atomically sharp despite some apparent La diffusion in SVO. Importantly, the interfaces 1 and 2 are chemically symmetric with a similar level of La/Sr intermixing. Quantitatively, the La signal intensity drops to its minimum within an atomic layer at both interfaces but is still present inside the SVO layer with an intensity of 25\% . To what extend this value represents the real La/Sr intermixing level will be discussed later on.

Fig.\ref{Vmap}g and Fig.\ref{Vmap}h show the spectral weight map of the \ce{V^{3+}} and \ce{V^{4+}} signal using multiple linear least-squares fitting with the reference fingerprints (Fig.\ref{Spectrum}) \cite{Tan2011}. From a qualitative point of view, the inspection of spectra recorded across the interface indicates that the \ce{V^{3+}} signal is mainly located in the LVO layers and the \ce{V^{4+}} signal is in the SVO layer as expected. The V oxidation state across the film (Fig.\ref{Vmap}b, gray) evolves from a minimal value of $\sim$3.1 in LVO to a maximum of $\sim$3.6 in SVO. The oxidation state averaged over a whole unit cell of the superlattice is +3.35(5) and corresponds to the expected value +3.33. The average values of +3.56(5) in the SVO layer and +3.24(5) in the LVO layer are quantitatively off the expected values of +4 and +3. This can be partly attributed to multiple elastic scattering or inelastic delocalization inherent to STEM-EELS experiments \cite{Dwyer2005}.

\begin{figure}
\centering
\includegraphics[width=0.85\linewidth]{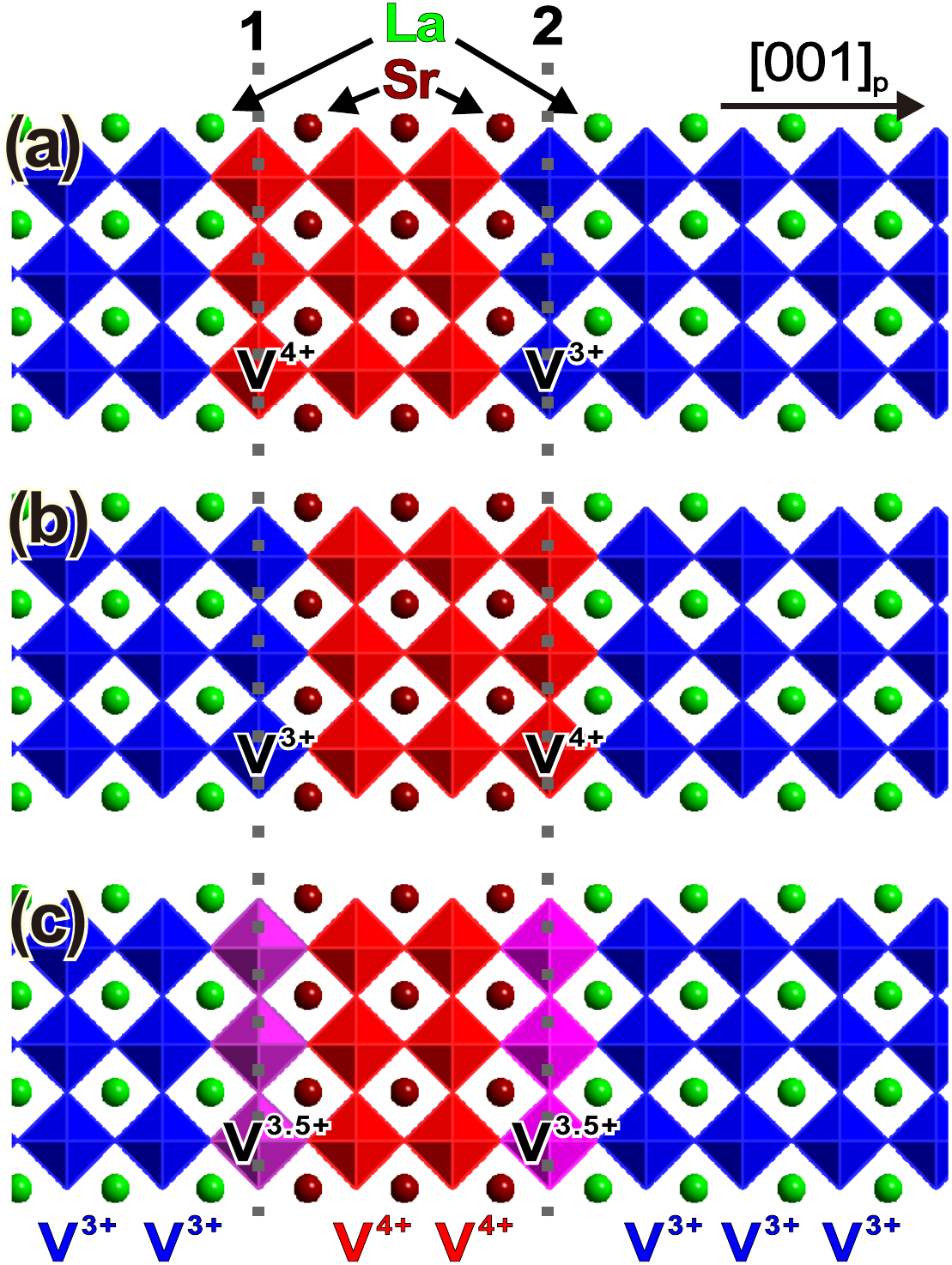}
\caption{(color online) Sketch of the simplified interface structure for three different situations (a) asymmetric case with 4+ (red octahedrons) and 3+ (blue octahedrons) and (b) with 3+ and 4+ at interface 1 and 2, respectively. (c) symmetric case with $V^{3.5+}$ on both interfaces (purple octahedrons). }
\label{Sketch}
\end{figure}

Both the \ce{VO_2} interfacial layers at interface 1 and 2 have a neutral SrO layer on one side and a positively charged \ce{LaO^+} layer on the other side (Fig.\ref{Vmap}de). However, the \ce{V^{3+}} and \ce{V^{4+}} signal maps show that the V oxidation state at interfaces 1 and 2 are different (Fig.\ref{Vmap}gh). Comparing this to the La elemental map, which precisely indicates the interfaces between the LVO and SVO layers, the \ce{V^{4+}} signal is shifted as a whole by half a unit cell against the growth direction (Fig.\ref{Vmap}h). The significant difference in oxidation state at both interfaces can be clearly observed in the \ce{V^{4+}} profile averaged along the interface direction (Fig.\ref{Vmap}c). The \ce{V^{4+}} profile peaking at interface 1 is approximately twice the intensity of that at interface 2 (Fig.\ref{Vmap}c). Quantitatively, the V valences of 3.53(4) and 3.38(4) measured at the interface 1 and 2 confirms this asymmetry with the difference being well above the error bars. Simplifying this observation to a model, we can assume interface 1 to contain \ce{V^{4+}} similar to the SVO layers and interface 2 to have a \ce{V^{3+}} state similar to the LVO layers (Fig.\ref{Sketch}a).

\begin{figure}
\centering
\includegraphics[width=0.95\linewidth]{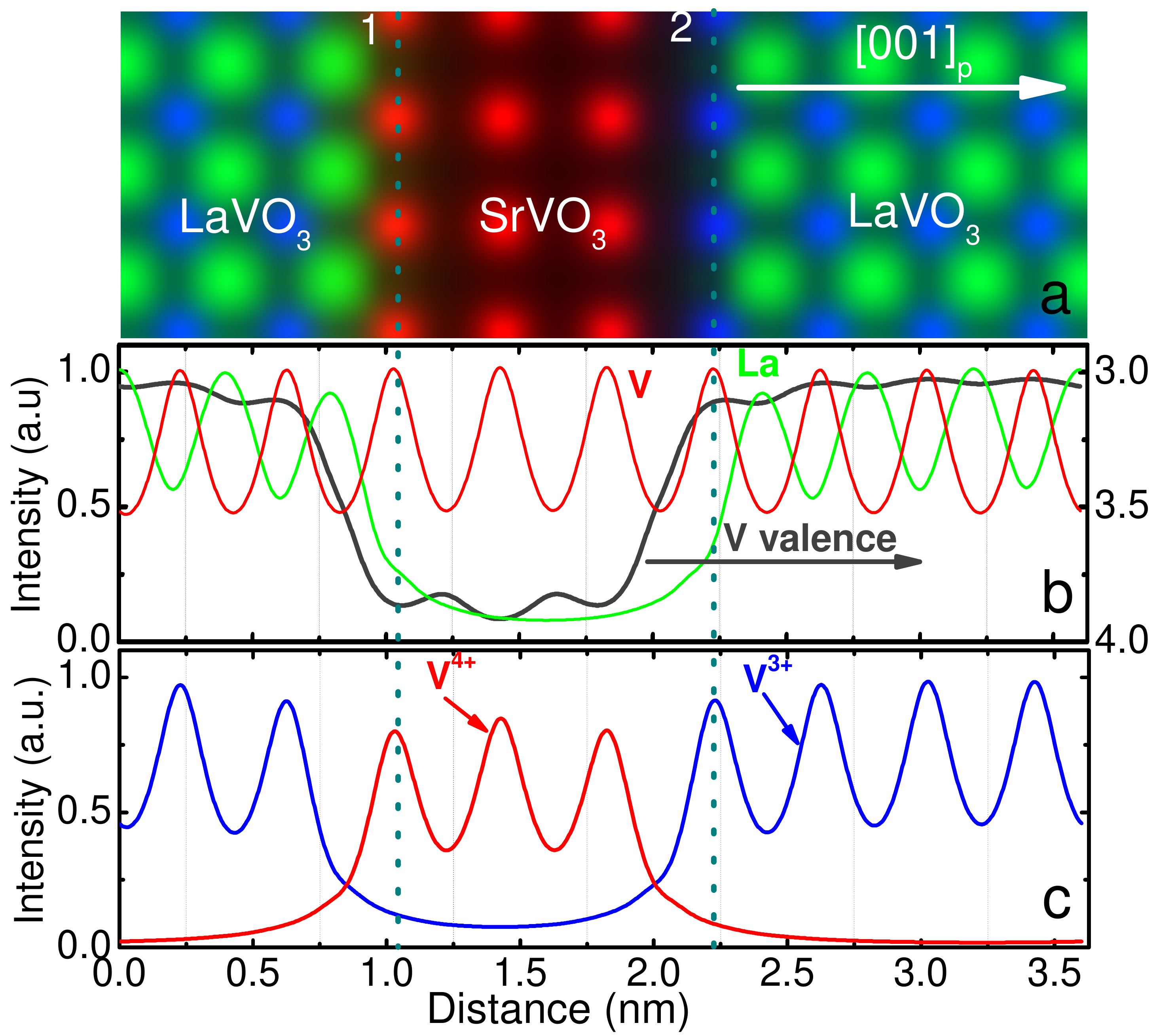}
\caption{(a) combined color map from the simulated maps of \ce{V^{4+}}(red), \ce{V^{3+}} (blue) and La (green), corresponding to the structure shown in Fig.\ref{Sketch}a. (b) intensity profile of V and La signal which can be compared to Fig.\ref{Vmap}d. The simulated V valence profile is shown in grey line. (c) intensity profiles of the \ce{V^{3+}} and \ce{V^{4+}} signal. The interface 1 is in 4+ state while the interface 2 is in 3+ state.}
\label{Simu}
\end{figure}

Considering the crystal structure shown in Fig.\ref{Sketch}a, STEMsim software was applied in order to do a double channeling inelastic simulation (Fig.\ref{Simu})\cite{Dwyer2010}. The combined simulated elemental map and profile is qualitatively comparable with our experimental result (Fig.\ref{Vmap}). After taking into account the elastic scattering, inelastic delocalization and source size broadening in the simulation, the La intensity drops to approximately 10\% in the SVO layer compared to 100\% in the LVO layer. In the experimental case the signal drops to approximately 25\%. The extra 15\% La signal in our experiment could therefore represent a rough estimate of the true La intermixing in the SVO layer. The \ce{V^{4+}} and \ce{V^{3+}} signal profiles agree well with the experiment with notably the \ce{V^{4+}} intensity profile largely symmetrical and consisting of three equally intense layers of \ce{V^{4+}}. This illustrates the general validity of the proposed interface structure (Fig.\ref{Sketch}a). Nevertheless, the tail of  \ce{V^{4+}} and \ce{V^{3+}} signal profiles at the interfaces extend further in the experiment than in the simulation. The simulation also reveals a more abrupt variation in the V oxidation state (Fig.\ref{Simu}b and Fig.\ref{Vmap}b). In Fig.\ref{Simu}b, the V oxidation changes from a minimal plateau at +3.07(1) in the LVO part to a maximal plateau at +3.85(3) in the SVO part. Experimentally, the V valence variation at the interface is much smoother as if some 4+ charge from the SVO layer ``leaks'' into the LVO layer (Fig.\ref{Vmap}b).

Such charge ``leakage'' at heterointerfaces was originally reported to be symmetric both in experimental results \cite{Ohtomo2002, Garcia2010} and in theoretical calculations \cite{Okamoto2004} (Fig.\ref{Sketch}c). However, at a relatively large scale (10~nm), Nakagawa et al. discovered that a 5~nm \ce{Ti^{3+}} layer is present at an \ce{AlO_2/LaO}/\ce{TiO_2} interface, while a similar layer is absent or much weaker at an \ce{AlO_2/SrO}/\ce{TiO_2} interface \cite{Nakagawa2006}. They also reported that the former interface has a roughness of 3~nm, twice as rough as the latter. Here we also observed similar asymmetry at an atomic sharp interfaces. A difference can be observed in the V oxidation state after applying our method to other regions with symmetric La chemical diffusion interfaces 1 and 2. Occasionally, situations as in Fig.\ref{Sketch}b and Fig.\ref{Sketch}c are observed.

The local strain asymmetry observed in Fig.1 is indeed the most likely candidate as a direct relation between bond length and valency exists \cite{Brown1985}. This strain mediated valence asymmetry is likely caused during the growth process with strain buildup that can not completely relax \cite{Gariglio2007}

Besides the unexpected asymmetric behavior of the V oxidation state at the two interfaces, the V oxidation state was found to be more delocalised than the La signal, indicating an electronic reconstruction with charge transfer up to 1.2~nm (Fig.\ref{Vmap}bc). This charge transfer distance is close to and may explain the proposed coherence length of the charge carrier in the \ce{(LVO)_m}/\ce{(SVO)_n} system, which is approximately between 1.3~nm to 2.2~nm\cite{Sheets2009}. This result is also equivalent to findings at interfaces between two insulating materials with a length scale for charge transfer typically about 1.2~nm \cite{Hwang2012,Cantoni2012, Okamoto2004}.

\section{Conclusion}

Using high spatial and energy resolution STEM-EELS, we quantitatively map the oxidation state of transition metal and the electronic reconstruction at metal/insulating interfaces. An electronic reconstruction was found up to 1.2~nm, likely related to the length of charge transfer in this system. Asymmetric electronic states are found at chemically symmetric interfaces and point to an issue unaddressed so far. This could be closely related to the asymmetric strain we observed at the interfaces. Importantly, this work provides evidence that the picture of interfacial polar discontinuities inducing electronic reconstruction should exist in many other systems than LAO/STO, including metal/insulating interfaces. To what extent the ``unexpected" properties discovered at TMO interfaces can be attributed to the existence of a so-called ``2D electron gas" is still the subject of debate. Such experiments are crucial and demonstrate the ability of aberration-corrected monochromated STEM-EELS to give a detailed insight in the atomic and electronic reconstructions at oxide interfaces.

\begin{acknowledgments}
 The Qu-Ant-EM microscope was funded by the Hercules foundation of the Flemish Government. Funding from the European Research Council under the 7th Framework Program (FP7), ERC grant N°246791 - COUNTATOMS and ERC Starting Grant N°278510 VORTEX is acknowledged. This work was partially funded by the European Union Council under the 7th Framework Program (FP7) grant N°NMP3-LA-2010-246102 IFOX, FWO Belgium and a GOA project ``XANES meets ELNES'' of the University of Antwerp. The authors acknowledge financial support from the European Union under the Seventh Framework Program under a contract for an Integrated Infrastructure Initiative. Reference No. 312483-ESTEEM2. The authors thank Prof. Martin Hytch for useful discussions.
\end{acknowledgments}

\end{document}